\documentclass[twocolumn,pra]{revtex4}
\begin{document}
{\noindent \bf Reply to ``Comment on
`Superfluidity in the interior-gap states' "}\\*[3mm]

%###################
In a recent paper, Liu and Wilczek\cite{LW} proposed the existence of a
uniform superfluid state (the interior-gap state) in systems of two species
Fermions with unequal masses and mismatched Fermi surfaces at weak coupling.
Following this proposal, we investigate in Ref.~\cite{WY} the stability of
this state by calculating its superfluid density and discover that at weak
coupling the superfluid density is always negative. We therefore conclude
that the spatially uniform (gapless) interior-gap state can not be stable at
weak coupling.

In response to our work, Liu and Wilczek criticized in a recent
Comment\cite{LW2} that our conclusion was in error since the interior-gap
state can be stable for systems with fixed particle densities. The only
part of this Comment\cite{LW2} directly relevant to our work is the paragraph
which contains their equation (2). Unfortunately, their argument in this
part is incorrect. Contrary to their claim, there can actually be no
linear corrections $\sim |{\bf w}|$ to the Fermi momenta $p_{F\alpha}$
when the particle densities are held fixed. In
fact, even if they existed, such corrections would yield no contributions to
the paramagnetic current due to
spatial isotropy. From their equation (2), it is indeed quite obvious that
there can only be corrections to the
paramagnetic densities which are higher order in ${\bf w}$. Fixing the
particle densities, therefore, cannot save the superfluid density
from turning negative. Thus the interior-gap state is unstable even under
this constraint.

This work was supported by the National Science Council of Taiwan
under Grant No.~91-2112-M-001-063.\\*[5mm]

{\noindent Shin-Tza Wu and Sungkit Yip\\
\hspace*{5mm}Institute of Physics\\
\hspace*{5mm}Academia Sinica\\
\hspace*{5mm}Nankang, Taipei 115, Taiwan}\\*[3mm]

{\noindent
Date \today \\
PACS numbers: 03.75.Kk, 05.30.Fk, 67.90.+z, 74.20.-z}

\end{document}